\documentclass[twocolumn,showpacs,preprintnumbers,amssymb]{revtex4}

\usepackage{graphicx,epsf, epsfig, amssymb}
\usepackage{bm}
\usepackage{longtable}
\usepackage{color}

\def\be{\begin{equation}}
\def\ee{\end{equation}}
\def\beq{\begin{eqnarray}}
\def\eeq{\end{eqnarray}}

\newcommand{\secintro}{I}
\newcommand{\secsetup}{II}
\newcommand{\secresults}{III}
\newcommand{\secconclusion}{IV}

\begin{document}

\title{Cross section, final spin and zoom-whirl behavior in high-energy black hole collisions}

\author{Ulrich Sperhake$^{1}
$, Vitor Cardoso$^{2,3}
$, Frans Pretorius$^{4}
$, Emanuele Berti$^{1,2}
$, Tanja Hinderer$^{1}
$, Nicolas Yunes$^{4}
$}
\affiliation{${^1}$ California Institute of Technology, Pasadena, CA 91109, USA}
\affiliation{${^2}$ Department of Physics and Astronomy, The University of
Mississippi, University, MS 38677, USA}
\affiliation{${^3}$ CENTRA,~Dept.~de F\'{\i}sica,~Inst.~Sup.~T\'ecnico,
  Av.~Rovisco Pais 1, 1049 Lisboa, Portugal}
\affiliation{${^4}$ Department of Physics, Princeton University, Princeton, NJ 08544, USA}
%


\begin{abstract}
We study the collision of two highly boosted equal-mass, nonrotating black
holes with generic impact parameter. We find such systems to exhibit
zoom-whirl behavior when fine tuning the impact parameter.  Near the threshold
of immediate merger the remnant black hole Kerr parameter can be near maximal
($a/M\gtrsim 0.95$) and the radiated energy can be as large as $35\pm 5 \%$ of
the center-of-mass energy.
\end{abstract}



\maketitle

\noindent{\bf{\em \secintro. Introduction.}}
As the major foundation of experimental particle physics, high energy
collisions enable us to probe the fundamental characteristics of short range
interactions: the larger the center-of-mass (CM) energy of the collision, the
shorter the distance one can probe. The mass-energy equivalence of relativity,
however, imposes a distinct limit on this method. All forms of energy
gravitate and collisions at trans-Planckian energies are expected to
generically form black holes (BHs). Attempts at probing shorter distances by
increasing the CM energy of the system are in consequence barred by the
formation of the event horizon \cite{Giddings:2002kt}.  Because the
gravitational field in this regime is predominantly sourced by the kinetic
energy, the nature of the interaction should be rather insensitive to the
internal structure of the particles \cite{Choptuik:2009ww}.
The trans-Planckian scattering of point
particles should therefore be well described by BH scattering
\cite{banks_fischler}.

Ultra-relativistic BH scattering simulations are relevant for recent proposals
to solve the hierarchy problem by adding ``large'' extra dimensions
\cite{Arkani-Hamed:1998rs} or an extra dimension with a {\em warp} factor
\cite{Randall:1999ee}, which allow for the Planck scale to be near the
electroweak scale. This offers the exciting possibility to produce BHs in
particle colliders and ultra high-energy cosmic ray interactions with the
atmosphere \cite{banks_fischler,bhprod}.  Classical BH production
cross sections and the energy radiated in gravitational waves (GWs) are inputs
for the Monte-Carlo simulators used to search for such events
\cite{Frost:2009cf}.  Further applications of high-speed BH collisions to
high-energy physics have recently been suggested by the AdS/CFT
correspondence~\cite{adscft,Nastase:2005rp}.

This {\em Letter} reports on several remarkable features arising in
ultra-relativistic BH collisions with generic impact parameter, thus
generalizing the head-on collision results of~\cite{Sperhake:2008ga}. We also
extend a recent preliminary exploration by Shibata {\it et al.}
\cite{Shibata:2008rq}, which estimated the impact parameter at the threshold
between BH formation and scattering for different boost magnitudes.  Unless
stated otherwise, we use geometrical units $G=c=1$.

\noindent{\bf{\em \secsetup. Setup.}}
%
The simulations presented in this work have been obtained with the {\sc Lean}
code \cite{Sperhake:2006cy}, which is based on the {\sc Cactus} computational
toolkit \cite{Cactusweb}. The code employs mesh refinement (provided by {\sc
  Carpet} \cite{Schnetter:2003rb}) and the apparent horizon (AH) finder {\sc
  AHFinderDirect} \cite{Thornburg:1995cp}. Puncture initial data are provided
by a spectral solver \cite{Ansorg:2004ds}. The simulations are performed as in
the head-on case \cite{Sperhake:2008ga}, but here we use equatorial instead of
octant symmetry and employ higher resolution near the BHs.

We set up a coordinate system such that the BHs start on the $x$-axis
separated by a coordinate distance $d$ and with radial (tangential) momentum
$P_x$ ($P_y$).  The impact parameter is $b\equiv L/P = P_y d/P$,where $P$ is
the linear momentum of either BH, and $L$ is the initial orbital angular
momentum. We extract gravitational radiation by computing the Newman-Penrose
scalar $\Psi_4$ at different radii $r_{\rm ex}$ from the center of the
collision.  $\Psi_4$ is decomposed into multipoles $\psi_{lm}$ using
spin-weight $-2$ spherical harmonics:
$\Psi_4(t,r_{\rm ex},\theta,\phi)=\sum_{l=2}^\infty
\sum_{m=-l}^l \,{_{-2}}Y_{lm}(\theta\,,\phi)\, \psi_{lm}(t,r_{\rm ex})$,
where $\theta$ is measured relative to the $z$-axis.  The estimated spurious
or ``junk'' radiation in the initial data is quite insensitive to the impact
parameter and comparable to that present in the head-on
case~\cite{Sperhake:2008ga}; we remove it from reported results in the same
manner.  Errors due to discretization and finite extraction radius are
comparable to those reported in \cite{Sperhake:2008ga}.  The estimated
uncertainties in radiated quantities are $3\%$ and $15\%$ for low and high
boost, respectively, and the phase error in the GW signal used in the
zoom-whirl analysis below is 0.2~rad.

Our analysis of grazing collisions is based on three one-parameter sequences
of numerical simulations of equal-mass, nonspinning BH binaries. Each sequence
is characterized by fixed initial coordinate separation $d$ and Lorentz boost
$\gamma\equiv(1-v^2)^{-1/2}$ of the holes, while the impact parameter $b$ is
varied. These sequences are
(1) $\gamma=1.520$ ($v = 0.753$) and $d/M=174.1$;
(2) $\gamma=1.520$ ($v = 0.753$) and $d/M=62.4$;
(3) $\gamma=2.933$ ($v = 0.940$) and $d/M=23.1$,
where $M$ is the total BH mass.

\noindent{\bf{\em \secresults. Results.}}
The results of our study are most conveniently presented in terms of three
distinct regimes we encounter as the impact parameter is increased starting
from the head-on limit $b=0$: (i) {\em immediate} mergers, (ii) {\em
  nonprompt} mergers and (iii) the {\em scattering} regime where no common AH
forms.  These regimes are separated by two special values of $b$: the {\em
  threshold of immediate merger} $b^{*}$ and the {\em scattering threshold}
$b_{\rm scat}$.  The remarkable features of BH binaries in these different
regimes will be described in detail in the remainder of this section.

The scattering threshold $b_{\rm scat}$ is defined such that the two BHs merge
for $b<b_{\rm scat}$ and scatter to infinity for $b>b_{\rm scat}$.  By
analyzing collisions with CM velocity $v\lesssim 0.90$, Shibata {\it et al.}
\cite{Shibata:2008rq} estimate $b_{\rm scat}/M\sim \left(2.5\pm
0.05\right)/v$.  The analysis of sequences 1 and 2 shows that merger occurs
only for $b_{\rm scat}/M\le 3.4$, consistent with \cite{Shibata:2008rq}, but
for sequence 3 we find $2.3 \lesssim b_{\rm scat}/M \lesssim 2.4$, indicating
that \cite{Shibata:2008rq} may overestimate $b_{\rm scat}$ for large $\gamma$.
Previous studies in the literature, which are based on the Penrose
construction, look for AHs in the union of two shock waves and find $b_{\rm
  scat}/M=1.685 M$ as $v\to 1$ \cite{Yoshino:2005hi}.  Our results suggest
that estimates of BH production cross sections ($\propto b_{\rm scat}^2$)
obtained through that construction are accurate to within a factor $<2$.  We
further emphasize the surprising agreement between our simulations and the
point-particle approximation: for example, a cross section estimate from
high-energy scattering off a Kerr BH with $j\simeq 0.98$ gives $b_{\rm scat}/M
\simeq 2.36$ \cite{Berti}.

%
The above definition of the scattering threshold is purely based on the nature
of the end state of the binary but ignores details of the interaction. A
closer look at these details reveals the existence of a {\em threshold of
  immediate merger} $b^*$ \cite{Pretorius:2007jn}. Roughly speaking, for
$b<b^*$ merger occurs within the first encounter, whereas for $b^*<b<b_{\rm
  scat}$ it does not, but sufficient energy is radiated to put the binary into
a bound state that {\em eventually} results in a merger.
A more precise definition arises in the context of the geodesic limit.
The argument is that this threshold
should generically be accompanied by behavior akin to {\em zoom-whirl} orbits
in the geodesic
limit~\cite{Pretorius:2007jn,Grossman:2008yk,Healy:2009zm}. For point
particles orbiting BHs, the existence of zoom-whirl orbits is intimately
related to that of {\em unstable} spherical orbits (at radii $3\leq r/M\leq 6$
for Schwarzschild BHs).  One unusual property of these orbits is that an
infinitesimal perturbation causes the particle to either fall into the BH or
``zoom'' out on a bound elliptic orbit (if initially $4< r/M< 6$ in the
Schwarzschild example) or along a hyperbolic orbit (if initially $3< r/M<
4$). Conversely, a one-parameter family of geodesics that smoothly
interpolates between capture by the BH ($b<b^*$) and not ($b>b^*$) will evolve
arbitrarily close to one of these unstable orbits as $b\to b^*$. The number of
``whirls'' close to the unstable orbit is given by
\be
\label{n_gamma} 
n = C - \Gamma \ln |b-b^*|\,, 
\ee
where $C$ is a family-dependent constant and $\Gamma$ is inversely
proportional to the Lyapunov instability exponent of the limiting spherical
orbit. Although unstable spherical orbits are a formal idealization, not
realized in practice because of GW energy loss, BH mergers do indeed approach
a whirl-like configuration near the threshold of immediate merger.  Such a
configuration can in principle be sustained until all the {\em excess} kinetic energy,
roughly equal to $2m_{irr}(\gamma-1)$, is lost in GWs, where 
$m_{irr}$ is the irreducible mass of each black hole.  This threshold is
blurred to a size $\delta b\sim e^{(C-n)/\Gamma}$ in parameter space.



\begin{figure}[ht]
  \centering
  \includegraphics[width=8.0cm,clip=true]{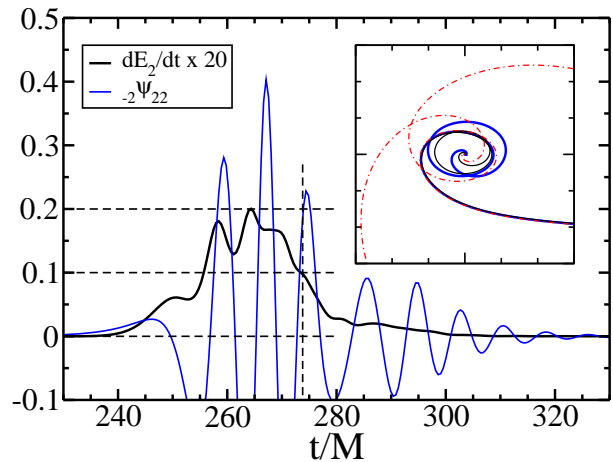}
  \caption{An illustration of how we estimate $n_{\rm GW}$, here for
    $b=3.36M$. The horizontal lines correspond to maximum, half-maximum and
    zero luminosity in the flux; the vertical line is when the flux drops
    below half maximum.  For reference, in the inset are puncture trajectories
    of one BH for this case (thick solid blue line), $b=3.34M$ (thin solid
    black line) and $b=3.39M$ (dash-dotted red line). All trajectories overlap
    early on during the whirl; the $b=3.39M$ trajectory peels off first before
    merging later, while the $b=3.34M$ case immediately spirals in after the
    whirl. The $b=3.36M$ case marginally peels off before merging, but it is
    clear from the luminosity that the merger/ringdown phase beginning after
    $t/M\sim 280$ is quite distinct from the preceeding whirl phase.}
  \label{fig:psi4r_22_de2all_dt}
\end{figure}

We now analyze sequence 1 in more detail to determine if it is consistent with
the zoom-whirl picture, and if an expression of the form of
Eq.~(\ref{n_gamma}) holds.  To this end we estimate the number of orbits $n$
in the whirl phase in two ways: (a) using the puncture trajectories ($n_p$),
and (b) using the GW flux measured far from the impact ($n_{\rm GW}$). For
method (a), in the scattering cases we define $n_p$ as the total angle divided
by $2\pi$ traversed by the puncture from the initial position until it reaches
a distance from the origin of twice the minimum distance (i.e., roughly twice
the whirl radius). For merger cases, $n_p$ is counted in a like manner until
the puncture crosses a distance $1/2$ the whirl radius.  Our current bracket
gives $b^*/M=3.35\pm0.01$, close to which we already see a ``blurring'' of the
threshold, as binaries with $b \gtrsim b^*$
do not separate to twice the whirl distance before merging (see the
inset of Fig.~\ref{fig:psi4r_22_de2all_dt}).  For method (b), we define
$n_{\rm GW}$ to be the number of GW cycles divided by $2$ in the $l=m=2$
component of the wave, from the initial time until the time
when the $l=2$ mode luminosity reaches $1/2$ its peak: see
Fig.~\ref{fig:psi4r_22_de2all_dt} for an example.  We expect this estimate
to be decent because the luminosity seems to be largest and
roughly constant during the whirl phase.  The ``$1/2$ criterion'' is somewhat
arbitrary, but as long as it is applied consistently it should have little
effect on our estimate of the {\em slope} $\Gamma$ in Eq.~(\ref{n_gamma}).

\begin{figure}[t]
  \centering
  \includegraphics[width=8.0cm,clip]{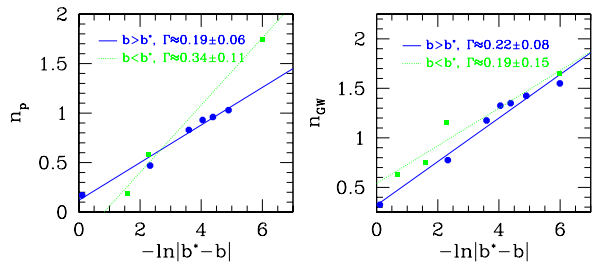}
  \caption{Estimated number of orbits using puncture trajectories (left) and
    GWs (right) as a function of distance from $b^{*}$ for immediate and
    nonprompt merger cases in sequence 1. The uncertainties in the fitted
    slopes to Eq.~(\ref{n_gamma}) are purely from the linear regression
    analysis.}
  \label{fig:crit_gamma}
\end{figure}

The two estimates $n_p$ and $n_{\rm GW}$ are shown in
Fig.~\ref{fig:crit_gamma} for immediate merger and nonprompt merger impact
parameters about $b^*$.  The plots indicate that a relationship of the form (\ref{n_gamma}) 
is valid, with a slope $\Gamma\approx 0.2$ to
within $\sim 50\%$ (ignoring systematic and computational errors we have not
been able to account for); this is a factor of 2-3 smaller than the analogous
geodesic problem of a high-speed point particle scattering (prograde) off a
Kerr BH with $j\approx 0.95$. Given the large range of relevant $\Gamma$'s in
the geodesic case (cf.~Fig.~9 of \cite{Pretorius:2007jn}) this provides
reasonable evidence for zoom-whirl-like behavior in high-speed,
comparable-mass collisions.

%

\begin{figure}
  \centering
  \includegraphics[width=8.5cm,clip]{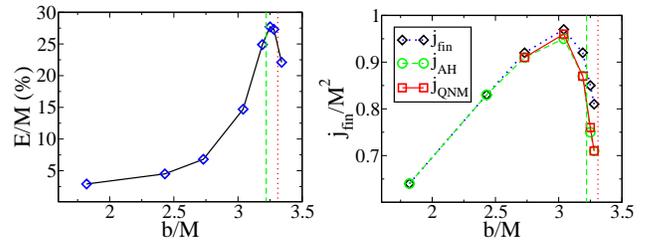}
  \caption{Total energy radiated (left) and final BH spin (right) vs. impact
    parameter from sequence 2, the latter calculated using several methods.
    The vertical dashed green (dotted red) line is the estimated immediate
    merger threshold $b^*$ (the scattering threshold $b_{\rm scat}$).
  }\label{fig:E_b}
\end{figure}

The threshold of immediate merger $b^*$ appears to play a special role when we
consider the amount of GW energy radiated and the final spin of the
post-merger BH. To illustrate this point, in Fig.~\ref{fig:E_b} we plot these
quantities as functions of the impact parameter for sequence 2.

For $v\sim 0.75$ the radiated energy increases by about one order of
magnitude, from $\sim 2.2\%$ for $b=0$ to $\gtrsim 23\%$ for $b\sim b^*$. Two
points are particularly noteworthy in this regard. (i) Even at this
comparatively small boost $v\sim 0.75$ we comfortably exceed the maximum of
$14\pm3~\%$ reported for the ultrarelativistic limit of head-on collisions
\cite{Sperhake:2008ga}.  Grazing collisions with larger boosts, in turn,
radiate enormous amounts of gravitational radiation: for one run of sequence 3
with $v\sim 0.94$ and $b\sim b^*$ the radiated energy is $\sim 35\pm5\%$ of
the CM energy. (ii) The maximum radiation as well as the maximum final spin
(cf.~below) is obtained near the threshold of immediate merger $b^*$ as
opposed to $b_{\rm scat}$.  The notion that excess kinetic energy drives
zoom-whirl behavior seems to be consistent with the data, in that the maximum
total energy radiated near $b^*$ is approximately equal to the initial kinetic
energy minus the spin energy of the final BH.

These surprisingly large amounts of GW energy correspond to huge luminosities.
Ref.~\cite{Sperhake:2008ga} showed that the high-energy, head-on collision of
two nonspinning BHs could generate luminosities up to $dE/dt\sim 0.01$. For
non head-on collisions and nonprompt mergers we observe even higher
luminosities.  For instance, for $v=0.75$ and $b\approx b^*$ the maximum
luminosity is $\sim 0.02$, and extrapolation to $v=1$ indicates that one might
reach luminosities $\gtrsim 0.1$, corresponding in physical units to $\sim
3.6\times10^{58}{\rm erg}\, {\rm s}^{-1}$.  This is the largest luminosity
from a BH merger known to date, approaching in order of magnitude the
universal limit $dE/dt \lesssim 1$ suggested by Dyson \cite{dyson}.

%

We conclude our analysis with a discussion of the final spin resulting from
the merger.  It has been argued in Washik {\em et al.} \cite{Washik:2008jr}
that ``no merger of equal-mass BHs can lead to a final BH with maximal spin
parameter $j_{\rm fin} \approx 1$'', as long as the BHs are nonspinning. The
maximum spin reported for the sequence studied in that work is $0.823$.
Larger final spins have been obtained from binary simulations with nonzero
initial spin.  Dain {\em et al.} \cite{Dain:2008ck} report the largest value
measured so far ($j_{\rm fin}=0.922$), although Fig.~3 in \cite{Healy:2009ir}
might imply an even larger final spin.  The latter work expresses doubts,
however, whether merging equal-mass BHs can ``produce a very-close-to-maximal
final BH''.
\begin{table}
\caption{Radiated energy, angular momentum and final spin for three
  representative sequence 2 runs.}
\begin{tabular}{rcccccccc}
$b/M$ & $E_{\rm rad}/M$  & $J_{\rm rad}/M^2$ & $J_{\rm rad}/J_{\rm ini}$  & $j_{\rm fin}$ & $j_{\rm QNM}$&$j_{AH}$ \\
\hline
0.00  &2.2\%   & 0.00  & 0.00   & 0.00  & 0.00 & 0.00 \\
2.74  &6.8\%    & 0.25  & 0.25   & 0.92  & 0.91 & 0.91 \\
3.04  &14.7\%  & 0.45  & 0.40   & 0.97  & 0.96 & 0.95 \\
\hline \hline
\end{tabular}
\label{tab:jfin2}
\end{table}
Our simulations suggest a different outcome for the merger of equal-mass,
nonspinning BH binaries.  In Table~\ref{tab:jfin2} we report the largest final
spin measured so far in any numerical BH merger simulation. We further
conjecture that even equal-mass, nonspinning binaries can result in a final
spin arbitrarily close to the Kerr limit $j=1$.  It is important in this
context to bear in mind the difficulties in measuring the final spin with high
accuracy (cf.~\cite{Marronetti:2007wz}).  These difficulties were our main
reason to generate sequence 2. The high oscillation frequency of the ringdown
signal as $j \rightarrow 1$ requires high resolution in the GW extraction
zone.  By using a smaller initial separation, sequence 2 enables us to meet
this requirement at tolerable computational cost.

We have checked our results by calculating the final spin in a variety of
ways: (i) we used energy balance arguments to find $j_{\rm fin} = J_{\rm
  fin}/M_{\rm fin}^2=(J_{\rm ini} - J_{\rm rad})/(M_{\rm ADM} - E_{\rm
  rad})^2$; (ii) we fit the quasinormal mode (QNM) frequency and damping time
of the final BH and inverted them to obtain $j_{\rm QNM}$ (see
e.g.~\cite{Berti:2009kk}); (iii) we used the equatorial circumference of a
Kerr BH $C_e=4\pi M$ to find $2\pi A_{AH}/C_e^2=1+\sqrt{1-j_{AH}^2}$, where
$A_{AH}$ is the AH area \cite{Kiuchi:2009jt}.  These different estimates are
compared in Table \ref{tab:jfin2} for three selected impact parameters leading
to merger, and shown in Fig.~\ref{fig:E_b} for sequence 2 runs.  Within our
uncertainty estimates ($\sim 3~\%$ for $j_{\rm QNM}$, $j_{\rm AH}$ and $\sim
8~\%$ for $j_{\rm fin}$) we observe good agreement throughout sequence 2.  For
$2.7\lesssim b/M\lesssim b^*/M$ we find $j_{\rm fin}>0.9$, and for $b \lesssim
b^*$ our estimated final spins can be quite close to extremality. For example,
for $b=3.04M$ we directly measure $j=0.96\pm 0.03$, and we expect further fine
tuning of $b$ to yield even larger values of $j$. Our estimates are
substantially larger than the ones quoted by Shibata {\it et al.}
\cite{Shibata:2008rq}.  Given the difficulties in achieving the necessary
numerical accuracy, perhaps the apparent discrepancies are due to our
increased resolution.

\noindent{\bf{\em \secconclusion. Conclusions.}}
%
High-energy BH collisions are fertile ground for testing many ideas and
conjectures in general relativity and high energy physics. We find that these
collisions can radiate at least $35\pm 5\%$ of the CM energy in GWs, that a
merger can lead to the remnant BH spinning very close to extremal, and we
display near-threshold phenomena akin to zoom-whirl in geodesics. We find no
evidence of cosmic censorship violation.  Our results are crucial for BH event
generators in TeV-scale gravity: for four-dimensional spacetimes our results
are consistent with Penrose's construction to within a factor $<2$. For the
first time we show the dependence of the spin of the BH remnant on the impact
parameter, a direct input in BH event generators.

{\bf \em Acknowledgements.}
This work was partially supported by FCT--Portugal through project
PTDC/FIS/64175/2006, NSF grants PHY-0745779, PHY-0601459, PHY-0652995,
PHY-090003 and PHY-0900735, the Alfred P. Sloan Foundation and the Sherman
Fairchild foundation to Caltech.  Computations were performed at TeraGrid in
Texas, Magerit in Barcelona, the Woodhen cluster at Princeton University and
HLRB2 Garching.


\end{document}